\documentclass[11pt, a4paper, onecolumn]{google}

\usepackage{longtable}
\usepackage{multirow}
\usepackage[square,numbers,sort&compress]{natbib}
\usepackage{amssymb}
\usepackage{float}
\usepackage{makecell}
\usepackage{graphicx}
\usepackage{caption}
\usepackage{afterpage}

\usepackage{xcolor}

\definecolor{linkblue}{RGB}{0,0,238}

\title{From Text to Discovery: How Large Language Models Are Reshaping Research Across Scientific and Humanistic Disciplines}

\renewcommand{\today}{}

\author[1]{Saleh Afroogh*}
\author[1]{Yasser Pouresmaeil, Yiming Xu}
\author[1]{Kevin Chen}
\author[1]{Abhejay Murali}
\author[1]{Junfeng Jiao}

\correspondingauthor{saleh.afroogh@utexas.edu}

\affil[1]{\small\textcolor{linkblue}{University of Texas at Austin}}

\begin{abstract}
Large Language Models (LLMs) are rapidly reshaping academic research across the natural sciences, social sciences, and humanities, yet the scientific community lacks a comprehensive, cross-disciplinary account of how these tools are being integrated, what they deliver, and where they fall short. This paper addresses that gap by mapping their current state and outlining an agenda for their responsible integration into scientific research.
Our analysis reveals a consistent pattern: LLMs meaningfully accelerate research workflows — from hypothesis generation and literature synthesis to data analysis and scientific writing — while introducing serious challenges related to hallucination, reproducibility, dataset bias, and model opacity. Beyond technical limitations, we identify ten underexplored challenges, including the erosion of researcher autonomy, AI-driven confirmation bias, authorship ambiguity, and unequal access to these technologies — systemic risks that demand interdisciplinary governance frameworks, robust validation standards, and expanded explainability research.\\
\hfill
\end{abstract}

\begin{document}

\maketitle

\section{Introduction}
Artificial intelligence (AI) has become increasingly integrated into scientific inquiry, transforming methodologies and practices across the humanities, social sciences, and natural sciences. Among the most significant recent developments are Large Language Models (LLMs)—like GPT-4 and Bard—which have emerged with sophisticated capacities for natural language comprehension and generation in natural language. Such models can help speed up research, increase its accuracy, and make it more accessible; they could also become powerful assistants for any text analysis or hypothesis generation in academic settings while fostering interdisciplinary collaborations.

Although the adoption of LLMs in academic contexts has been increasing, no comprehensive review to date has evaluated their use across scientific disciplines. Existing studies predominantly focus on specific applications, such as their role in accelerating discoveries in biology and chemistry \cite{1}, or their utility in assisting with academic writing and publication processes \cite{2}. However, these works do not address the broader challenges or cross-disciplinary implications of LLM integration. The identified gap highlights the need for a comprehensive review investigating not only the potential benefits but also the ethical dilemmas related to the use of large language models in diverse research settings.

This paper tries to fill this gap by reflecting on challenges and opportunities linked with using general-purpose LLMs, such as GPT-4,and specialized models across the humanities, social sciences, and natural sciences. Particular attention is given to ethical issues, including biases in generated content, intellectual property disputes, and potential over-reliance on AI systems. The analysis takes a cross-disciplinary approach, identifying recurring themes and issues that transcend disciplinary boundaries, while avoiding extensive focus on field-specific nuances.

To maintain a focused and rigorous analysis, this review concentrates on the humanities, social sciences, and natural sciences, where text-based methodologies are foundational to research practices. The exclusion of formal sciences (e.g., mathematics, computer science) and applied sciences (e.g., engineering, agriculture) is intentional, as these fields often emphasize algorithmic development, technical implementation, or creative applications that diverge from the textual and interpretive challenges examined here. By narrowing the scope, this review seeks to uncover trends and challenges more universally relevant to the selected disciplines while avoiding dilution of the analysis with already well-documented field-specific issues.

This study is guided by the need to balance the transformative potential of LLMs with the ethical and practical challenges they introduce. In doing so, it provides a foundation for informed, responsible integration of LLMs into scientific research, with implications for both current and future practices across disciplines.

\afterpage{%
    \begin{figure}[H]
        \centering
        \makebox[\textwidth][c]{%
            \includegraphics[
                width=1\textwidth,
                height=0.9\textheight,
                keepaspectratio
            ]{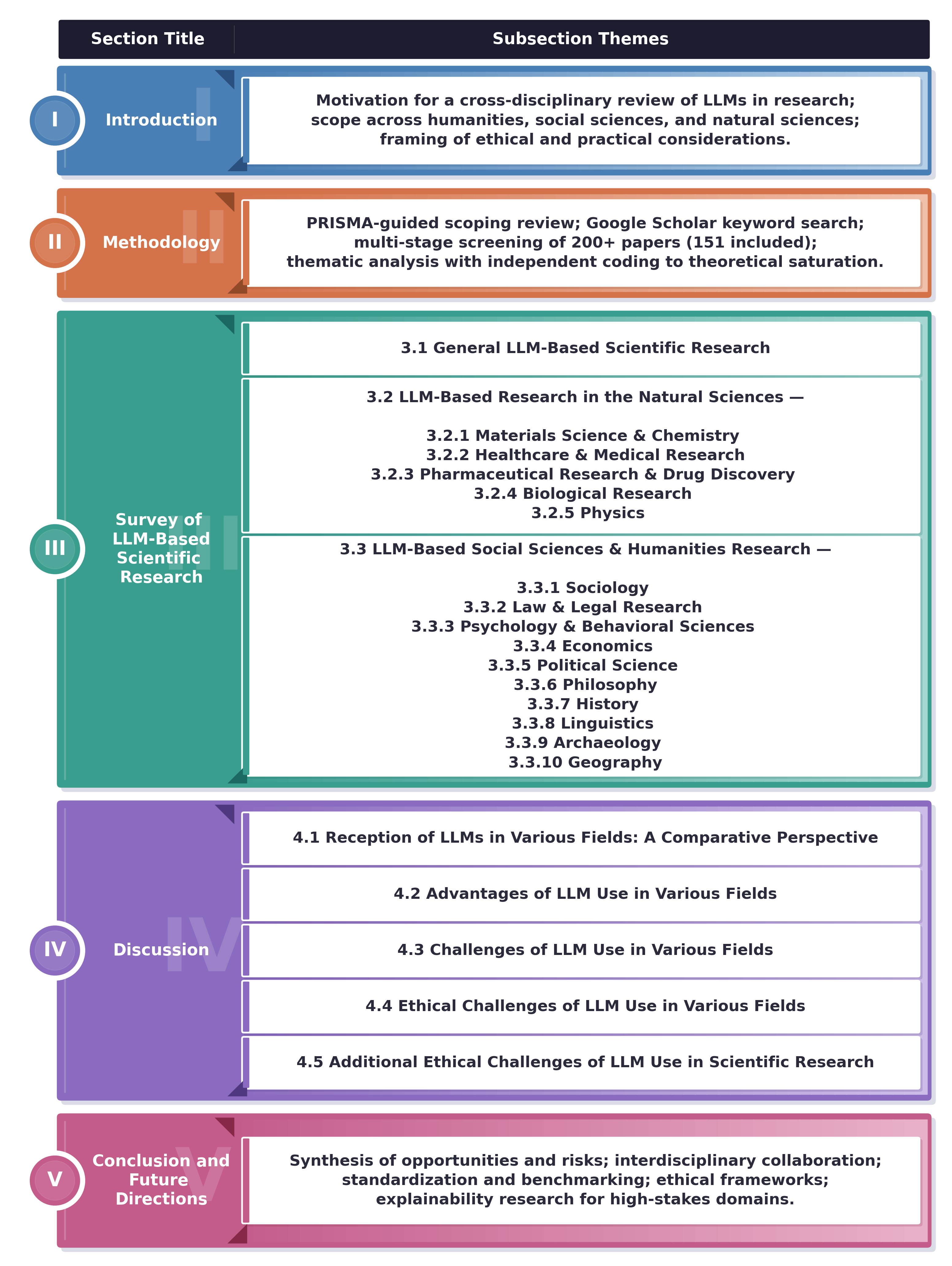}%
        }
        \caption{Schematic Overview of the Study}
        \label{fig:roadmap_table}
    \end{figure}
    \clearpage
}

\section{Methodology}
\label{subsec:2}
We undertake a systematic scoping review to synthesize and map the existing literature on the use of LLMs in different academic disciplines and professional fields. Scoping review methods are particularly well suited to the investigation of diverse research areas by identifying existing gaps in knowledge and mapping the central concepts of a field. This methodology was adapted to our purposes while remaining within the methodological framework set by the Preferred Reporting Items for Systematic Reviews and Meta-Analyses, PRISMA.

We conducted a search on Google Scholar using a systematic set of keywords to capture the breadth of research on LLM applications. The search strings included: "large language model use + scientific research," "LLM use + humanities research," "LLM use + social science research," and combinations of LLM with specific fields such as medical research, healthcare research, neuroscience, biology, sociology, and legal research.

The search generated an initial corpus of over 200 academic papers. A multi-stage screening process was then employed to ensure the inclusion of high-quality and relevant literature. All retrieved titles and abstracts were first reviewed to identify studies directly relevant to the application of LLMs in their respective fields. Papers were included when they were published in peer-reviewed journals, conferences, or credible databases and explicitly addressed the use or impact of LLMs within a specific discipline. Studies were excluded when they focused solely on technical details of LLM architecture without contextual application or when they were published in a language other than English.

Following the initial screening, 151 papers were deemed eligible and included in the final review. Key metadata and content from each paper were extracted and organized using a structured framework. This framework captured the field of study, such as social sciences, healthcare, and legal research, the type of LLM application, such as content generation, data analysis, and hypothesis testing, the research methodology employed, whether qualitative, quantitative, or mixed-methods, and the reported benefits, limitations, and ethical considerations.

The extracted data were synthesized using thematic analysis. Common themes and trends were identified and categorized to give an essential overview of the landscape of research. In order to minimize bias and increase the reliability of the study, we performed screening and coding independently. Any discrepancies were resolved through consensus. The iterative process of review and coding assured theoretical saturation, that is, that no new themes emerged in subsequent analyses.

Findings in this review provide a strong foundation for understanding the state of LLM applications in the disciplines, and more importantly, common themes and trends were identified and categorized to give an essential overview of the landscape of research.

\begin{figure}[H]
    \centering
    \includegraphics[
        width=0.9\textwidth,
        height=0.5\textheight,
    ]{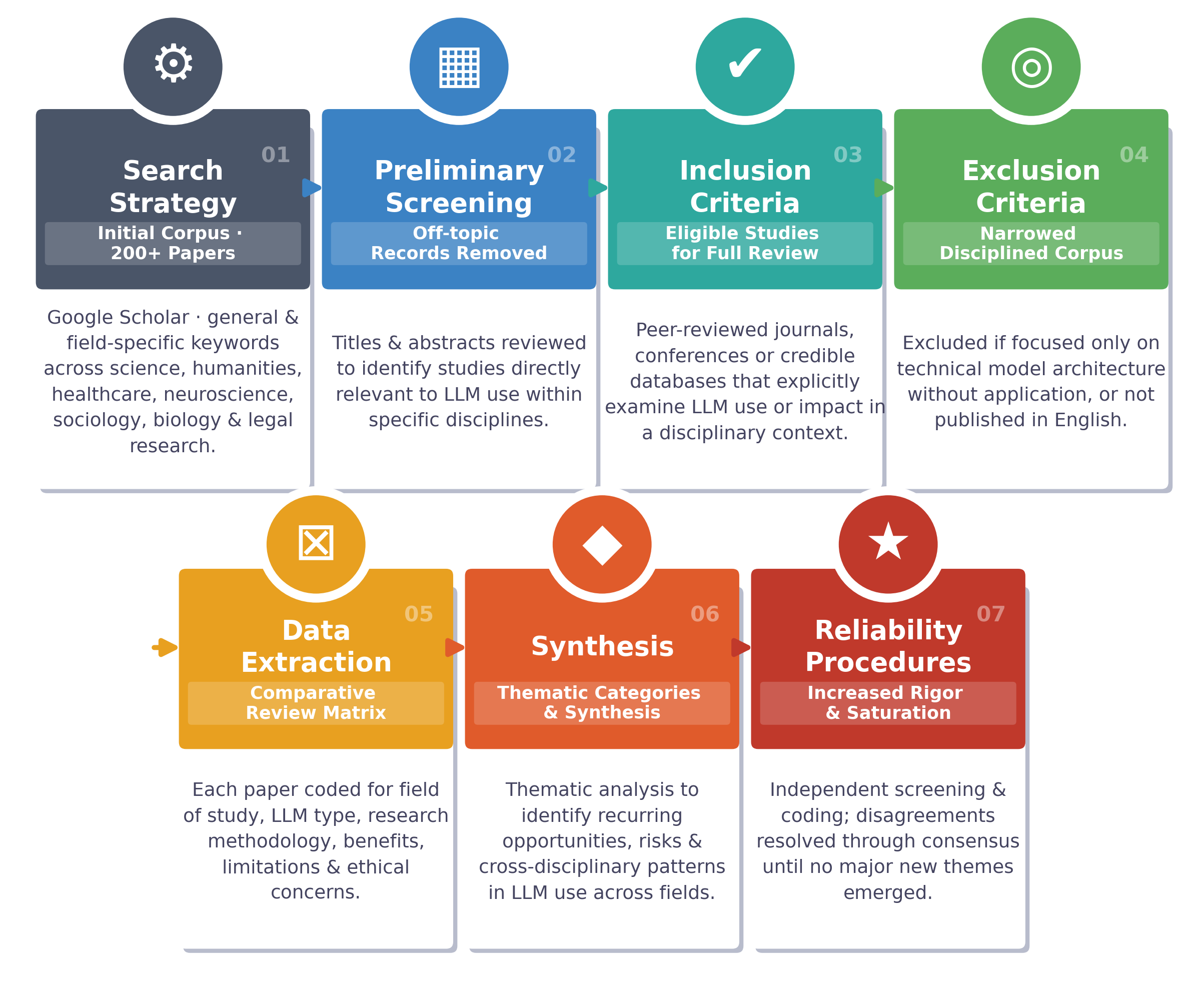}
    \caption{Review Design, Screening Logic, and Analytical Workflow}
    \label{fig:workflow_row}
\end{figure}

In order to minimize bias and increase the reliability of the study, authors performed screening and coding independently. Any discrepancies were resolved through consensus. The iterative process of review and coding assured theoretical saturation—that is, that no new themes emerged in subsequent analyses.

Findings in this review provide a strong foundation for understanding the state of LLM applications in the disciplines, and more importantly, they provide ample leads and justifications to spur further exploration and development.

\section{Survey of LLM-Based Scientific Research}

This section surveys the growing body of literature on LLM applications across scientific disciplines, organized into three broad domains: the natural sciences, the social sciences, and the humanities. Each domain presents a distinct set of opportunities and challenges shaped by its methodological traditions, data characteristics, and epistemological commitments. The survey begins with a cross-disciplinary overview of general LLM capabilities and limitations in research contexts, before turning to field-specific examinations that illuminate how these tools are being adopted, adapted, and critically assessed within individual disciplines. 

\subsection{General LLM-based scientific research}
LLMs have emerged as transformative tools in scientific research, offering a wide array of capabilities that streamline traditional practices and enhance various stages of the research process \cite{3, 4}. Their contributions extend across several areas of scientific discovery. They can help identify promising research questions by analyzing large datasets, revealing overlooked trends, and synthesizing substantial volumes of information \cite{5}. They also strengthen literature review practices by rapidly condensing extensive bodies of scholarship, allowing researchers to detect central findings, unresolved gaps, and possible future directions more efficiently \cite{6, 7, 8}. In programming-related tasks, LLMs assist with code generation, syntax correction, debugging, and general technical guidance, which can reduce development time and lower the frequency of coding errors \cite{7, 9, 10}. They are likewise useful in scientific writing, where they can organize material, draft text, address writer’s block, and support proofreading, thereby helping researchers produce more polished manuscripts \cite{5, 7, 10, 11}. Beyond writing support, LLMs contribute to complex data analysis, especially in tasks involving natural language processing and pattern recognition, making the examination of large datasets more efficient \cite{11, 12}. They may also provide feedback on manuscripts and assist with idea generation, although the originality and practicality of these suggestions can vary \cite{13,14,15,16,17,18}. In addition, LLMs can synthesize knowledge across sources and explain complicated concepts in ways that support knowledge translation \cite{19}. Some scholars have even suggested that they may enable new and more accessible forms of scientific dissemination beyond conventional peer-reviewed publishing \cite{20}.

Despite their potential benefits, the widespread adoption of LLMs also raises several challenges that need careful consideration \cite{21}. One major issue involves dataset bias and reliability, since these models are trained on large-scale corpora that may embed distortions, inaccuracies, or other biases, which can contribute to hallucinated or otherwise unreliable outputs \cite{5,7,9,10,11,22}. Ethical concerns are also significant, particularly with respect to plagiarism and intellectual property, as questions remain about the originality of AI-generated material when it draws on preexisting works \cite{5,9,11}. Citation integrity presents another challenge, given that LLMs may generate inaccurate references or even cite non-existent studies, thereby threatening academic rigor \cite{11,23}. Questions of authorship and accountability further complicate their use, especially when LLMs play a meaningful role in shaping research ideas or producing written content \cite{5,11,24,25}. Finally, access to advanced models may remain uneven, and the growing reliance on paid systems could deepen disparities between researchers in high-income and low-income settings, limiting equitable access to these technologies \cite{5}.

To address these challenges, human oversight is essential to ensure the reliability of LLM-generated content and prevent it from replacing critical human judgment \cite{5,11}. Recent studies have proposed guidelines for the ethical and effective use of LLMs, focusing on transparency, accountability, and fairness \cite{26}. However, debates about the feasibility of fully autonomous AI-driven research that adheres to scientific standards are ongoing \cite{27,28}.

The growing integration of LLMs into research workflows is akin to the historical adoption of other disruptive technologies in science. While resistance or excessive regulation may slow progress, embracing LLMs and developing innovative strategies to address their challenges will enable researchers to maximize their potential. The consistent availability, speed, and supportive nature of LLMs make them valuable collaborators in scientific endeavors. By carefully considering the ethical implications and ensuring proper oversight, the scientific community can responsibly harness the potential of LLMs to advance research.

\subsection{LLM-Based Research in the Natural Sciences}
Recent studies have highlighted several vulnerabilities inherent in applying LLMs to natural science research. Key challenges include factual inaccuracies, susceptibility to user prompt injection attacks (where malicious prompts exploit LLM vulnerabilities), and limitations in reasoning abilities. Furthermore, LLMs often lack up-to-date knowledge, fail to recognize potential risks associated with long-term planning, and may inefficiently allocate resources or become trapped in feedback loops. Additionally, handling multi-purpose or multi-tool tasks remains problematic, as does ensuring effective oversight of tool usage. Concerns also include the absence of specialized expertise, which may result in safety-critical lapses, insufficient or low-quality human feedback, inadequate environmental feedback, and reliance on unreliable research sources. These issues can compromise both the safety and efficacy of LLM applications in scientific contexts \cite{29}.

LLMs function as sequence predictors—assembling strings of vectorized symbols based on statistical patterns in prior text—but they inherently lack the embodied and relational capacities required to create scientific meaning. Unlike human scientists, who engage collaboratively in constructing and sustaining shared experiential realities, LLMs operate without participation in the lived contexts of scientific inquiry. This limitation undermines their ability to grasp nuanced value judgments, such as uncertainties, limitations, and complexities that are critical in scientific writing. As a result, relying solely on LLMs for tasks like scientific summarization risks producing oversimplified or potentially misleading interpretations of research findings \cite{21}.

\begin{figure}[H]
    \centering
    \includegraphics[
        width=1.0\textwidth,         
        height=0.88\textheight,      
        keepaspectratio
    ]{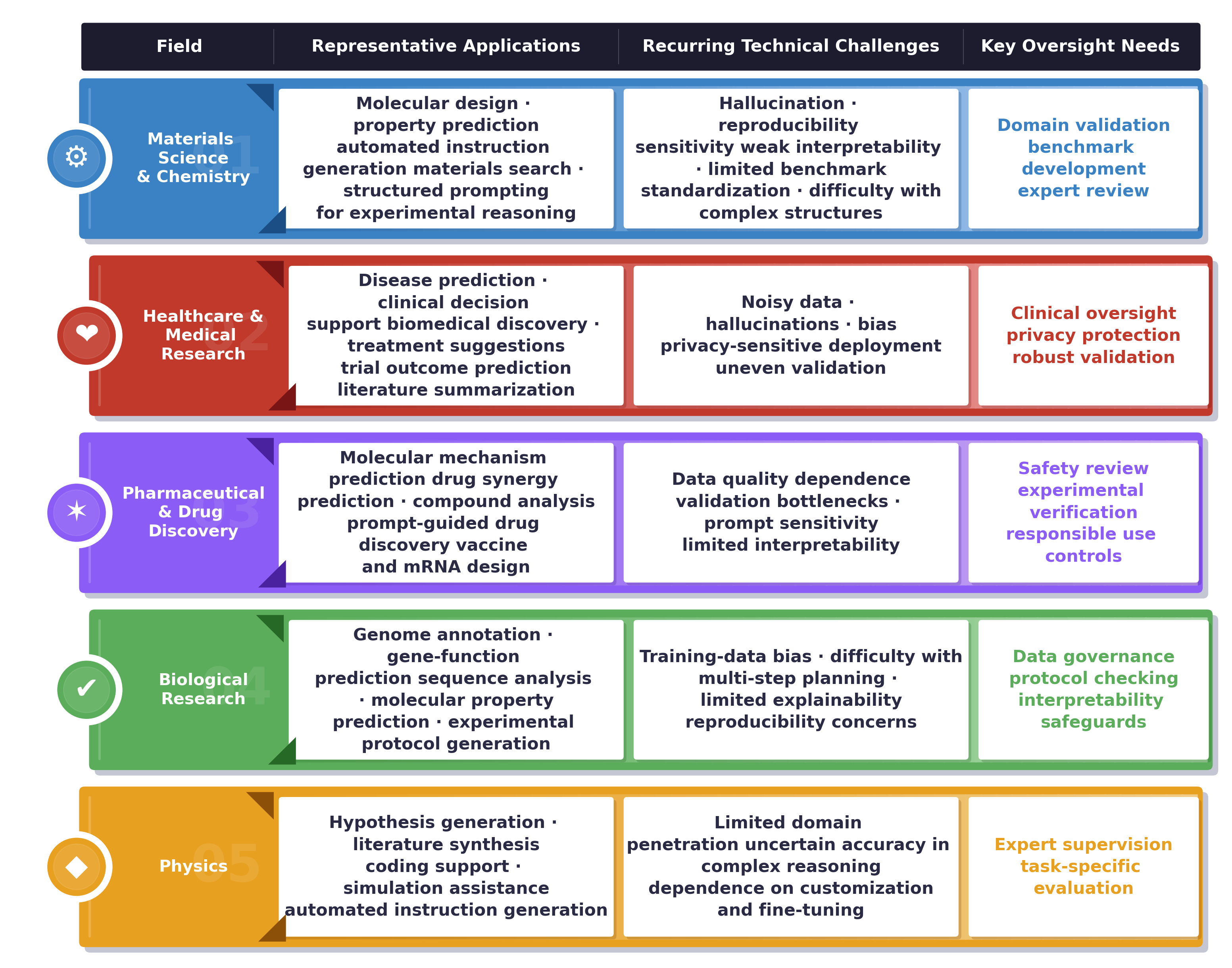}
    \caption{LLM Applications, Challenges, and Oversight Needs in
             Natural Science Research.}
    \label{fig:table1_natural_science}
\end{figure}

Further challenges arise in specialized fields such as drug discovery, where the absence of domain-specific expertise in LLMs can lead to significant lapses in safety, especially in high-stakes environments. The potential for LLMs to misinterpret complex scientific data or overlook critical safety protocols remains a substantial barrier to their widespread use in these areas.

Despite these limitations, LLMs show promise in accelerating certain aspects of scientific research, such as literature reviews, data processing, and hypothesis generation. However, their deployment in safety-critical areas must be approached with caution, ensuring rigorous oversight and ongoing refinement of their models. The following sections will explore how LLMs are applied across various natural science disciplines, identifying both the contributions and challenges unique to each field.

\subsubsection{Materials Science and Chemistry}
The application of LLMs to materials science and chemistry represents one of the most active frontiers in AI-assisted natural science research. These models have demonstrated notable capacity for accelerating core research tasks, including molecular design, property prediction, and synthesis optimization, while also enabling new forms of knowledge extraction from unstructured scientific literature. At the same time, their integration into these domains is not without difficulty, as concerns related to data quality, model interpretability, hallucination, and reproducibility continue to pose significant obstacles. The following subsections examine the current state of LLM deployment in materials science and chemistry, considering both the advances achieved through fine-tuning and structured prompting and the persistent limitations that must be addressed to ensure reliable scientific application.

\textbf{Comprehensive Reviews:} More recent studies have discussed the potential of LLMs in transforming materials science and chemistry, with particular application to tasks of molecular design, prediction of properties, and optimization of processes for synthesis. However, several challenges continue to impede their effectiveness: data quality and model interpretability are recognized as significant issues, but there is a lack of standard benchmarks for their development and application \cite{30}. While LLMs can extract valuable insights from chemical and materials science data, their application requires careful integration of domain expertise to guide and validate their outputs \cite{31}.

\textbf{Fine-Tuned Models and Applications:}
Fine-tuning—creating domain-specific models out of general-purpose ones—holds promise as one of the most effective ways to bring greater accuracy and relevance to LLMs in these areas. The fine-tuned LLMs show increased performance in tasks such as material property prediction and knowledge extraction from unstructured chemical data. For example, fine-tuning has enabled LLMs to generate automated instructions from scientific texts, streamlining workflows in chemical research \cite{32}. Additionally, LLMs have been applied to predict the physical and electronic properties of materials based on textual descriptions, demonstrating their potential to advance research in materials science \cite{33,34}.

Several studies have compared the performance of fine-tuned LLMs with traditional machine learning models, showing that LLMs often outperform these methods in chemical research tasks. For example, GPT-3, when fine-tuned to answer chemical questions in natural language, delivered results with significantly higher accuracy than baseline methods, highlighting the potential of these models for enhancing research productivity \cite{35,36}. Furthermore, a benchmark dataset named TextEdge has been developed to assess the performance of LLMs in predicting material properties from crystal structure descriptions, showcasing the adaptability and utility of LLMs in materials science \cite{37}. Despite these advancements, challenges remain in using LLMs for complex computational tasks, such as generating highly accurate crystal structures or predicting material behavior under varied conditions. To mitigate these issues, integrating LLMs with specialized knowledge bases and computational tools has been proposed to improve the reasoning and accuracy of predictions \cite{38}. 

Structured prompting techniques, such as the ordered and structured chain-of-thought (OSCoT) method, have been shown to significantly enhance GPT-4’s performance in predicting experimental outcomes and providing intuitive reasoning for its predictions \cite{39}. Another fine-tuned LLM system demonstrated high accuracy in searching, predicting, and generating materials with user-specified properties based on natural language input \cite{40}. Nonetheless, concerns persist regarding the robustness and reliability of LLMs for widespread application in materials science, emphasizing the need for further improvements \cite{41}.

\textbf{Hallucination and Reproducibility Challenges:}
Despite the promising applications of LLMs in materials science, significant challenges remain, particularly concerning hallucination and reproducibility. Hallucination usually refers to "the tendency of LLMs to generate false" or inaccurate information, which can be especially problematic when the models are tasked with complex topics like predicting material properties. For example, GPT-4 has been observed to make errors when generating silicon crystal structures, providing incorrect atomic positions that were only partially corrected after additional prompting \cite{42}. These models also tend to produce generic or modal responses, which may perpetuate existing biases. Similarly, GPT-4’s predictions for the electrical conductivity of inorganic materials were only marginally better than random guessing and included several misclassifications, which underscores the challenge of ensuring accurate and reliable outputs in scientific contexts.

Reproducibility, another major concern, can be significantly affected by the way prompts are phrased. Slight variations in input can lead to substantial differences in output, which is problematic in scientific research, where reproducibility is a cornerstone. This issue arises because LLMs, as sequence predictors, are heavily reliant on statistical patterns from training data and may fail to generate consistent results when faced with similar yet subtly different queries. This lack of reproducibility not only affects the reliability of LLM-generated content but also complicates the ability to verify and reproduce results, a core aspect of scientific inquiry.

\textbf{Limitations of LLM Application in Chemistry and Materials Sciences:}
LLMs are believed to exhibit several limitations, such as a tendency to memorize rather than truly understand and challenges in transferring implicit knowledge across semantically similar inputs. To address these issues, some studies propose a comprehensive benchmark dataset designed to assess LLM performance in predicting molecular properties \cite{43}.

In summary, LLMs offer significant potential to revolutionize research in materials science and chemistry by enhancing productivity, improving data extraction, and accelerating scientific discovery. However, challenges related to hallucination, reproducibility, and bias must be addressed before LLMs can be reliably integrated into research workflows. The ongoing development of fine-tuning techniques, improved prompting methods, and the incorporation of domain-specific knowledge bases are crucial steps in overcoming these challenges. As LLMs continue to evolve, their role in advancing materials science and chemistry will likely become more pronounced, but careful oversight and further research will be necessary to ensure that their application adheres to the highest standards of scientific rigor and ethical responsibility.

\subsubsection{Healthcare and Medical Research}
The integration of LLMs into healthcare and medical research has emerged as one of the most consequential and rapidly evolving areas of AI application in the natural sciences. These models hold considerable promise for transforming a broad range of research and clinical functions—from disease prediction and biomedical discovery to clinical decision support and drug development—by enabling the processing of large and heterogeneous datasets at a scale previously unattainable. However, the high-stakes nature of medical contexts also amplifies the importance of addressing key concerns, including data quality, patient privacy, model reliability, and ethical accountability. The following sections explore the specific applications, research efficiency gains, and ethical challenges associated with LLM deployment in healthcare and medical research, offering a comprehensive view of both the opportunities and the risks that accompany their growing adoption in this field.

\textbf{Overview and Current Landscape:} The application of LLMs in the fields of medicine and healthcare has drawn significant attention. Detailed surveys provide an overview of the global scenario of LLM applications in health science literature and give suggestions for future applications in academic and clinical practice \cite{44,45,46,47,48}. LLMs have demonstrated their ability to process diverse scientific data types, including publications, patents, textbooks, electronic medical records, DNA and protein sequence databases, and chemical compound libraries. When carefully validated, such systems can reason across modalities and maximize their utility in these domains \cite{49}.

\textbf{Applications of LLMs in Healthcare and Medical Research:}
LLMs have been applied across a diverse and expanding range of healthcare and medical research tasks, reflecting the breadth of their analytical capabilities and their potential to support both clinical and investigative work. From early disease detection and biomedical discovery to clinical decision support and trial outcome prediction, these models are increasingly being integrated into research workflows and patient care processes alike. The following subsections survey the principal areas of application, highlighting the specific contributions of LLMs and the contexts in which they have demonstrated the greatest utility.

\textit{Disease Prediction and Progression Analysis:} 
LLMs have been deployed to enhance early disease detection and progression analysis. For instance, models have been used to predict diseases such as sepsis \cite{50}, track the progression of chronic conditions \cite{51,52}, and make diagnostic predictions \cite{53}. Additionally, wearable sensor data has been integrated with LLMs to provide insights into various health conditions, including cardiovascular risks and metabolic disorders \cite{54}. During pandemics, LLMs have proven valuable in forecasting severity and mortality, demonstrating their potential in public health planning \cite{55}.

\textit{Biomedical Discovery and Domain-Specific Applications:}
In biomedical research, domain-specific fine-tuned LLMs are being applied to healthcare challenges and scientific discovery. These models have facilitated advancements in understanding protein-protein interactions (PPIs), predicting protein properties, and unraveling the molecular mechanisms of diseases \cite{56,57}. By processing vast datasets, these models accelerate the discovery of treatment targets and the creation of new diagnostic methods \cite{58,59,60}.

\textit{Clinical Decision Support:}
LLMs have emerged as a transformative tool in clinical decision-making. For example, in dental medicine, they support tasks such as summarizing patient data, assisting multilingual communication, and streamlining clinical documentation. In surgical science, LLMs are utilized for text generation, patient management, and surgical data extraction \cite{61,62}. While these applications improve efficiency, concerns surrounding patient confidentiality, data security, and ethical usage must be addressed.

\textit{Treatment Suggestions:}
LLMs assist in generating treatment recommendations for complex medical cases, such as those involving antibiotic-resistant infections \cite{63}. They also streamline drug discovery and development processes by analyzing datasets, identifying potential drug candidates, and optimizing experimental designs \cite{64}.

\textit{Observational Studies and Causal Inference:}
In medical studies relying on real-world clinical data, LLMs have been proposed as tools to overcome challenges related to causal inference. By reducing dependence on large interdisciplinary teams, LLMs offer a cost-effective solution for conducting observational studies and deriving actionable insights \cite{65}.

\textit{Trial Outcome Prediction:}
Accurately predicting clinical trial outcomes is vital for resource allocation and portfolio prioritization in pharmaceutical research. LLMs have demonstrated their ability to enhance trial planning and improve decision-making \cite{66,67}.

\textit{Neuroscience and Cognitive Studies:}
In neuroscience, LLMs contribute by enriching datasets with meta-information, summarizing extensive research, and bridging gaps between siloed research communities. They enable the integration of noisy but interrelated findings to predict experimental outcomes, often surpassing human experts in certain benchmarks \cite{68,69}. Additionally, LLMs help identify cognitive concepts that better explain brain phenomena, opening new avenues for interdisciplinary research.

\textbf{Advancing Research Efficiency:}
Beyond their direct clinical and biomedical applications, LLMs have also demonstrated significant value in enhancing the broader efficiency of medical research processes. By supporting tasks such as literature synthesis and scientific writing, these models help researchers manage the growing volume of health-related knowledge and accelerate the dissemination of findings across both general and specialized contexts. The subsections below examine these contributions in greater detail, illustrating how LLMs are reshaping the infrastructure of medical inquiry as much as its content.

\textit{Literature Summarization and Analysis:}
The exponential growth of health-related publications presents challenges for researchers in keeping up with emerging knowledge. LLMs mitigate this by summarizing large volumes of literature, identifying key insights, and streamlining systematic reviews \cite{70}. By generating code and supporting experimental design, LLMs save time and promote equitable access to research tools, enhancing adaptability in the scientific community \cite{64}.

\textit{Scientific Writing:}
LLMs also enhance productivity in scientific writing. By automating repetitive tasks, drafting documents, and proofreading, they reduce the workload on researchers and accelerate the dissemination of findings in both general and domain-specific contexts \cite{64}.

\textbf{Challenges and Ethical Considerations:}
Despite their considerable promise, the deployment of LLMs in healthcare and medical research is accompanied by a set of challenges that must be carefully managed. The sensitivity of medical data, the precision required in clinical contexts, and the potential for model-generated errors to carry serious real-world consequences together create a demanding environment for responsible AI integration. The subsections below outline the principal technical and ethical concerns identified in the literature, underscoring the need for robust validation, transparency, and ongoing human oversight in any LLM-assisted medical application.

\textit{Data Noise and Hallucination Risks:}
Healthcare data is inherently noisy, and the probabilistic nature of LLM outputs can lead to hallucinations or confabulations. These inaccuracies pose risks in clinical applications where precision is paramount \cite{58}.

\textit{Privacy, Bias, and Security:}
The use of LLMs in healthcare raises significant concerns about privacy, particularly regarding the leakage of sensitive patient data. Biases inherited from training datasets can lead to inequitable outcomes, and transparency issues in how models generate outputs remain unresolved \cite{58,64}. Additionally, cybersecurity threats, such as unauthorized access to healthcare systems, add another layer of risk. The potential for infodemics—misinformation amplification—and disputes over authorship and copyright in LLM-generated content further complicate their integration \cite{64,71,72}.

Overall, Overall, LLMs in medicine and healthcare hold considerable potential for advancing research, improving diagnostics, and supporting clinical decision-making. In order for the integration of LLMs to be effective, several critical challenges that need to be met: data quality, privacy, ethical concerns, refinement of models, increasing transparency, and rigorous validation processes that ensure robust results for the field. This will demand continued interdisciplinary collaboration, so the full potential of LLMs can be realized while risks are avoided and societal benefits are maximized.

\subsubsection{Pharmaceutical Research and Drug Discovery}
The rise of AI has significantly accelerated progress in drug discovery. Among the most recent revolutionary technologies, LLMs exemplify tools that enable unprecedented efficiency and capabilities across various phases of drug discovery, and reducing both the costs and delays associated with traditional methods \cite{73}.

\textbf{Computational Advancements in Drug Discovery:}
Integration of AI and deep learning (DL) in pharma research has redefined how biotech companies approach molecular modeling, opening truly new business models and research methodologies. LLMs like ChatGPT have now fast-tracked this change by enabling tasks that would have been very labor-intensive or even impossible to perform. These models are now employed to streamline the discovery of therapeutic molecules, promising substantial benefits for patients and reducing time-to-market for novel drugs \cite{73,74,75,76,77,78}.

\textbf{Mechanistic and Predictive Applications of LLMs:}
Accurately predicting the mechanisms and properties of potential drug molecules is a critical step in the drug discovery pipeline. LLMs excel in this area by processing vast datasets to recognize patterns and predict outcomes with remarkable precision. This capability expedites the identification of promising candidates while reducing the costs and time traditionally associated with research and development \cite{79}. Additionally, LLMs’ ability to analyze sequential biological data, such as protein structures and gene expression, enhances their utility in uncovering complex molecular relationships crucial for drug efficacy and safety.

\textbf{Specialized Applications in Drug Design:}
LLMs have shown exceptional utility in specialized drug discovery tasks. They have been applied to predict the synergistic effects of drug combinations, particularly in rare tissues where structured data remain limited \cite{80,81}. They also assist in the analysis of molecular structures by identifying relationships between form and function that are essential for therapeutic development \cite{82,83,84}. Together, these applications highlight the potential of LLMs to address highly specific and technically complex problems in drug design.

\textbf{Optimization through Prompt Engineering:}
The effectiveness of LLMs in drug discovery is significantly enhanced through prompt engineering. Tailored prompts allow researchers to direct models toward specific targets or hypotheses, improving both the relevance and accuracy of insights generated. This customization ensures LLMs remain focused on solving domain-specific problems with maximum efficiency \cite{85}.

\textbf{LLMs in Clinical Pharmacy:}
Other than drug discovery, LLMs are playing a central role in clinical pharmacy. These models analyze and interpret complicated datasets to assist clinical decision-making, optimize the process of patient care, and improve therapeutic interventions. Thus, synthesizing diverse data inputs, LLMs give clinicians actionable insights to improve efficacy and safety within the dispensing of treatment plans \cite{86}.

\textbf{LLMs in Vaccine Discovery:}
Domain-specific LLMs have emerged as powerful tools in vaccine development, particularly for mRNA-based technologies. These models predict mRNA properties critical for designing effective vaccines and therapeutics. By analyzing large datasets, LLMs can identify key patterns and relationships that optimize the structure and function of mRNA sequences, accelerating the development of vaccines for emerging diseases \cite{87,88,89}.

\subsubsection{Biological Research}
LLMs are reshaping biological research and provide us with the ability to understand complex biological systems holistically, enabling new methodological possibilities. They augment analysis of large and diverse datasets, producing insights previously difficult to obtain with traditional techniques.

\textbf{Transformative Applications of LLMs in Biology:}
LLMs have demonstrated remarkable potential in advancing various aspects of biological research. They support genome annotation with high accuracy, help predict gene functions, and identify previously unrecognized relationships within genetic data \cite{90}. By analyzing DNA, protein, and gene expression data, these models can uncover complex patterns that deepen understanding of cellular systems and contribute meaningfully to genomics, proteomics, and transcriptomics \cite{91,92,93}. LLMs have also been used to address challenging problems in computational biology, particularly in molecular property prediction and protein structure prediction, with important implications for drug development, bioengineering, and the design of new therapeutics \cite{94,95,96,97,98}. These applications underscore the versatility of LLMs as tools for both foundational and applied biological research.

\textbf{Automation and Experimental Planning:}
One of the most promising applications of LLMs in biology is the automation of experimental protocols. By synthesizing information from diverse sources, these models can generate accurate and detailed experimental procedures, thereby streamlining research workflows and reducing the likelihood of human error \cite{99}. At the same time, they continue to face difficulties in managing complex multi-step experimental plans, prompting researchers to develop automatic evaluation frameworks aimed at improving the accuracy and reliability of generated protocols \cite{99}. Early comparisons between LLMs and human experts suggest that these systems may become valuable research assistants, particularly in automating routine tasks and supporting hypothesis generation \cite{100}.

\textbf{Ethical and Technical Challenges:}
While LLMs offer transformative potential, several challenges must be addressed to maximize their utility in biological research. Biases embedded in training data can produce skewed or inaccurate predictions, especially for biological entities that are underrepresented in existing datasets. Protecting sensitive biological information and patient data also remains a major concern. In addition, the black-box nature of LLMs makes their outputs difficult to interpret, raising persistent concerns about transparency and reproducibility. Broader ethical questions related to equitable access and responsible use must likewise be approached with care \cite{101}. Addressing these obstacles proactively is essential if the full potential of LLMs in molecular biology and related fields is to be realized.

\subsubsection{Physics}
LLMs are emerging as powerful tools in physics, although they remain less widely used, offering support across various domains, including astronomy. They can support ideation and hypothesis generation, helping researchers explore new questions and stimulate scientific creativity \cite{102}. They also assist with literature review and coding by synthesizing complex scientific materials and generating code for simulations or data analysis, thereby improving both productivity and accessibility in research workflows \cite{103}. In addition, fine-tuned LLMs have been proposed as a means of improving research accuracy, with automated instruction generation from scientific texts representing one further direction for innovation \cite{32}.

\subsection{LLM-Based Social Sciences and Humanities Research}
The integration of Artificial Intelligence, particularly Large Language Models (LLMs), into the social sciences and humanities, is now fundamentally reshaping traditional research paradigms. These technologies hold substantial transformative potential by overcoming long-standing limitations, increasing productivity in research, and augmenting cognition. However, their adoption raises important methodological, ethical, and philosophical questions about the nature of scholarly work and the evolving boundary between human- and machine-generated contributions.

\begin{figure}[H]
    \centering
    \includegraphics[
        width=1.0\textwidth,         
        height=0.85\textheight,      
        keepaspectratio,
        trim=0 673pt 0 0,  
        clip
    ]{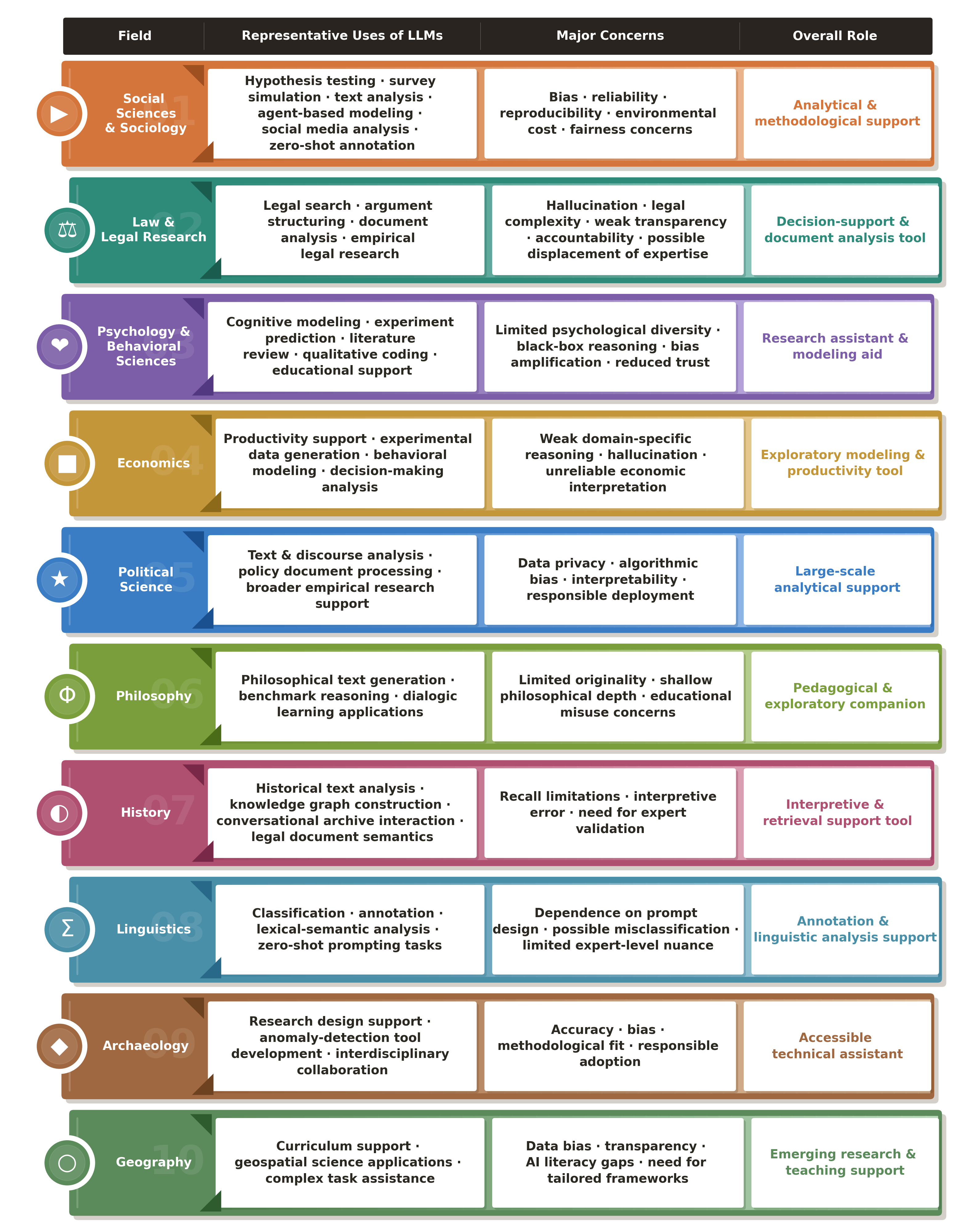}
    \caption{LLM Uses and Major Concerns Across Social Sciences
             and Humanities Fields.}
    \label{fig:table2_ssh_p1}
\end{figure}

\clearpage

\textbf{The Role of LLMs in Humanities and Social Sciences Methodology:}
LLMs are increasingly recognized as valuable tools for supporting a wide range of research methodologies in the humanities and social sciences. They can assist qualitative inquiry by helping researchers structure academic writing, formulate interview questions, and generate counterexamples that challenge existing interpretations, thereby supporting the iterative character of qualitative research and opening new directions for inquiry and hypothesis formation \cite{104,105}. They also facilitate textual analysis by streamlining tasks such as annotation, sentiment analysis, classification, and critical discourse analysis, which are especially important in fields such as sociology, political science, and literary studies, where close attention to language and meaning is essential \cite{106}. In deductive coding, for instance, LLMs can apply predefined codebooks to textual samples and systematically identify themes or codes in qualitative datasets \cite{107}. In addition, by simplifying the organization and analysis of large and complex datasets, LLMs can reduce the time and resource constraints that often limit ambitious research projects, while improving the scalability and breadth of qualitative studies. These applications demonstrate how LLMs can extend traditional research methods by enhancing analytical depth, efficiency, and precision.

\textbf{Challenges and Concerns in LLM Application:}
Despite their promise, the use of LLMs in humanities and social sciences research is accompanied by substantial challenges. One major concern is their tendency to produce highly confident responses even when the information is inaccurate or incomplete, a phenomenon commonly described as hallucination, which creates clear risks in research settings where accuracy is essential \cite{108}. LLMs may also reproduce and amplify the dominant narratives and biases embedded in their training data, which is especially problematic in qualitative research that seeks to recover diverse perspectives and question the prevailing social assumptions; unlike human scholars, these systems lack the capacity for reflexive judgment and ethical self-examination \cite{108}. Furthermore, although LLMs are effective in recombining existing knowledge and producing coherent summaries, they remain limited in generating genuinely original insights, since the production of new knowledge often depends on creativity, critical reflection, and engagement with human experience in ways that current models do not fully replicate \cite{109}. Their use in scholarly work also raises broader concerns about ethical and academic responsibility, particularly in sensitive research areas, while challenging conventional understandings of authorship and accountability.

\textbf{Collapsing the Humanities-Science Divide:}
The growing adoption of LLMs in the humanities and social sciences challenges the long-standing distinction between these disciplines and the sciences. The humanities have traditionally emphasized qualitative and interpretive forms of inquiry, while the sciences have generally favored quantitative and objective investigation. However, LLMs, through their ability to generate complex forms of text such as poetry, narrative, and analytical prose, increasingly blur the boundary between these traditions and invite renewed reflection on the intellectual methods that define them \cite{110}. At the same time, by producing language that can closely resemble human authorship, LLMs challenge the assumption that creativity is an exclusive human capacity, raising broader philosophical questions about the meaning of originality and the evolving role of the researcher \cite{110}. Even so, these systems need not be understood as replacements for human scholars. Rather, they may be viewed as collaborators that can augment human creativity and intellectual labor, encouraging a more integrative model of knowledge production in which AI functions as a complementary tool.

Overall, the integration of LLMs into social sciences and humanities research represents both an opportunity and a challenge. While these models offer powerful tools for analysis, efficiency, and innovation, their adoption necessitates careful consideration of ethical implications, methodological rigor, and the evolving definition of intellectual work. As researchers navigate this new landscape, LLMs may not only transform research processes but also deepen our understanding of the intersection between human and machine intelligence.

\subsubsection{Social Sciences and Sociology}
The advent of Large Language Models (LLMs) has significantly transformed research methodologies in social sciences and sociology. These models offer innovative approaches for simulating human behavior, testing theoretical frameworks, and modeling complex social dynamics. By enabling researchers to automate hypothesis generation and simulate social phenomena, LLMs expand the scope and depth of social scientific inquiries \cite{111,112}.

\textbf{Applications of LLMs in Sociology and Social Sciences:}
LLMs have introduced a wide range of applications in sociology and the broader social sciences. They enhance hypothesis testing and data analysis by streamlining conventional research workflows. By drawing on structural causal models, researchers can design experiments, formulate research questions, and analyze outcomes more efficiently, while LLMs further accelerate these processes through automated hypothesis generation and testing, thereby expanding the pace and scale of sociological inquiry \cite{113}. In addition, these models support several important methodologies. They can function as virtual respondents in survey research and online experiments, enabling more efficient data collection and interpretation \cite{114,115}. They are also effective in content and text-based analyses, including sentiment analysis, discourse analysis, and annotation, often reducing the need for extensive manual coding \cite{116,117}. LLMs have likewise shown value in agent-based modeling, where they simulate complex social interactions and help researchers examine collective behavior and decision-making processes \cite{118}. Their usefulness extends further to the simulation of digital and social media interactions, where they can model user opinions and communication patterns, offering insight into public sentiment and the spread of information in real time \cite{119}. Within computational social science, LLMs have also proven effective as zero-shot annotators, producing high-quality annotations and explanations without substantial fine-tuning, which is especially beneficial in settings with limited labeled data \cite{120}. They are also becoming increasingly capable of predicting the outcomes of social science studies, including inter-attitude correlations, through approaches such as fine-tuning and optimized prompting, thereby strengthening methodological rigor and supporting theory development \cite{121}.

\textbf{Challenges and Ethical Considerations:}
Despite their considerable promise, the integration of LLMs into sociology and the social sciences raises a number of important concerns. Bias and fairness remain central issues, as the datasets used to train these models frequently reflect existing social biases, which may in turn produce distorted or misleading research outcomes \cite{116}. There is also concern that these systems may exacerbate social inequalities, particularly when used in contexts requiring cultural sensitivity or involving marginalized communities \cite{114}. Questions of reliability and validity further complicate their use, since the possibility of producing low-quality or weakly grounded research through LLMs underscores the need for careful evaluation of generated outputs. Environmental sustainability has also emerged as a concern, given the substantial computational resources required to train and deploy large-scale models, prompting calls for more efficient approaches \cite{114}. In addition, issues of replication and reproducibility persist, as differences in datasets, model versions, and parameter settings can make it difficult to reproduce findings consistently across studies.

\textbf{Toward "Augmented Sociology":}
In spite of these limitations, there is growing recognition of LLMs as useful tools for enabling a more collaborative form of sociological research, often described as augmented sociology \cite{122}. This perspective emphasizes the role of LLMs as ideation partners that can introduce fresh perspectives, bridge disciplinary divides, and support methodological innovation. By automating time-consuming and resource-intensive tasks, they can also improve scalability and efficiency, allowing researchers to concentrate more fully on interpretation and higher-level analysis. More broadly, the integration of LLMs may expand the range of questions that sociologists are able to investigate, opening new possibilities for studying complex social phenomena, from digital interactions to structural inequality. Overall, LLMs have contributed to a meaningful shift in social science and sociological research by offering powerful analytical tools while also challenging established methods. Their capacity to simulate, predict, and analyze social behavior creates important opportunities for inquiry, but addressing persistent concerns related to bias, ethics, and sustainability will be essential if they are to be used responsibly and effectively. As the field advances, LLMs are likely to serve not as replacements for human scholars, but as collaborators that enhance creativity, precision, and research impact.

In summary, LLMs have introduced a paradigm shift in social science and sociology, providing powerful tools for research while challenging traditional methodologies. Their ability to simulate, predict, and analyze complex social phenomena offers unprecedented opportunities for inquiry. However, addressing their limitations—bias, ethical concerns, and sustainability—will be critical to ensuring their responsible and effective use. As the field continues to evolve, LLMs are poised to enhance sociological research, not as replacements for human scholars, but as collaborators that amplify creativity, precision, and impact.

\subsubsection{Law and Legal Research}
Integration of LLMs into the traditional body of legal research transforms established methods, enhances efficiency, increases precision, and strengthens analysis. While recognized for their ability to aid legal practice, LLMs demonstrate use across tasks like legal search, argument structuring, document analysis, and empirical inquiry. While promising, these advancements are accompanied by challenges that warrant further exploration.

\textbf{Applications of LLMs in Legal Research:}
LLMs excel in improving legal search by leveraging supervised learning methods to identify factually similar cases. This capability allows legal professionals to quickly locate relevant precedents and statutes, streamlining case law research—a cornerstone of legal practice. In legal argumentation, LLMs can also support the interpretation of complex legal theories and assist in identifying key facts while helping formulate arguments that are consistent with established precedents and broader public policy considerations. By expanding the analytical tools available to legal practitioners, these systems can contribute to the development of more persuasive and well-reasoned legal arguments \cite{123,124}.

\textit{Legal Document Analysis:}
LLMs offer significant potential in analyzing and interpreting legal documents, including contracts, statutes, and judicial opinions. They can evaluate legal writing in terms of clarity, coherence, and persuasiveness, while also identifying weaknesses in argumentation or inconsistencies in legal reasoning. In this way, LLMs can strengthen the quality and rigor of legal discourse and assist professionals in drafting more effective legal documents \cite{125}.

\textit{Empirical Legal Research:}
LLMs have become more widely adopted in empirical legal studies, as they automate the extraction of structured data from unstructured legal texts and allow large-scale analyses of judicial decisions, legal trends, and policy impacts by researchers. The automation of such processes cuts down time and effort in conducting extensive legal analyses but has also opened up new avenues for quantitative and qualitative insights \cite{125}.

\textbf{Challenges and Limitations:}
Despite their promise, the application of LLMs in legal research faces several challenges. The complexity of legal systems presents a major obstacle, as legal language is highly nuanced, judicial practices vary across jurisdictions, and precedents evolve over time, all of which complicate model accuracy and adaptability \cite{126}. The reliability of LLM outputs is also closely dependent on the quality of the data on which they are trained, meaning that incomplete, inaccurate, or biased datasets can weaken the integrity of legal research. Moreover, although LLMs have demonstrated notable capabilities, including success on examinations such as the U.S. bar exam, ongoing problems related to factual inaccuracy and hallucination, as well as limited interpretability, raise serious concerns in high-stakes legal contexts \cite{127}. The black-box nature of these models further limits trust, since legal professionals may be unable to determine how particular outputs were produced, which is a serious concern in a field where precision and accountability are essential. In addition, the use of LLMs in legal research raises broader ethical and practical questions involving bias, fairness, and the possibility that automation may displace forms of human expertise. For this reason, the effective adaptation of AI tools to legal practice will require continued collaboration between technologists and legal professionals.

In summary, while LLMs hold considerable promise in transforming legal research, their application remains hindered by certain technical and practical limitations. Future research is needed to address these challenges, particularly in improving model transparency, data quality, and alignment with legal practices.

\subsubsection{Psychology and Behavioral Sciences}
Large Language Models (LLMs) are emerging as transformative tools in psychology and behavioral science, offering novel approaches to cognitive modeling, experimental prediction, and methodological enhancement. However, their integration into these fields necessitates careful consideration of their capabilities, limitations, and ethical implications.

\textbf{Cognitive Modeling and Psychological Mechanisms:}
LLMs demonstrate significant potential for understanding and modeling aspects of human cognition. Research indicates their impressive performance in reasoning and problem-solving tasks—central to cognitive psychology—where they often match or surpass the abilities of neurotypical adults in specific contexts. The associationist framework, which highlights the importance of long-distance associations in thought processes, provides a useful lens for understanding LLMs’ reasoning capabilities.

Despite these strengths, LLMs continue to face important limitations in areas such as causal cognition and planning, where the flexibility and foresight characteristic of human reasoning remain difficult to reproduce. Their reasoning abilities can nevertheless be improved through strategies such as providing additional examples, which function much like scaffolding in human learning, and expanding model parameters to strengthen contextual understanding. In this sense, certain forms of improvement in LLM performance parallel mechanisms of cognitive facilitation observed in human reasoning.

Interestingly, analyzing LLM errors offers insights into human cognitive biases, as these mistakes often mirror patterns of bias and heuristic-driven errors in human reasoning. This positions LLMs as valuable tools for cognitive modeling, with the potential to advance our understanding of intelligence and reasoning mechanisms \cite{128}.

\textbf{Predictive and Experimental Applications:}
LLMs have shown considerable promise in predicting the outcomes of complex behavioral science experiments. For instance, studies indicate that GPT-4 performs at a level comparable to human experts when forecasting findings from experiments involving emotions, gender, and social perception \cite{129}. These predictive capabilities may assist researchers in generating hypotheses and in evaluating experimental designs before investing in costly empirical studies. Such developments suggest that LLMs may become increasingly important in experimental psychology by improving both the efficiency and the accuracy of research workflows.

\textbf{Enhancing Research Methodologies:}
LLMs have also demonstrated value in streamlining systematic literature reviews and qualitative coding. In one study, ChatGPT-4-turbo-preview was assessed on its ability to code 39 scientific papers, and its results were found to be comparable to those produced through traditional human coding \cite{130}. In these contexts, LLMs offer clear advantages in terms of cost-effectiveness by reducing the time and labor required for demanding research tasks, while also improving scalability through their capacity to process large datasets efficiently. At the same time, human oversight remains essential to ensure the validity and reliability of findings. Careful prompt design and rigorous evaluation are therefore necessary to maximize the usefulness of these systems while limiting the risks of bias or misinterpretation.

\textbf{Educational and Practical Applications:}
The capacity of LLMs to generate human-like language makes them useful in educational and pedagogical settings. They can assist with academic writing and can also be used to simulate aspects of human behavior for training and assessment purposes. However, caution is required when extending these applications into practical or clinical contexts. Because LLMs do not fully replicate human cognition or emotional understanding, they remain limited in tasks that demand deep empathy or complex social judgment \cite{131}. For this reason, their incorporation into educational and applied settings must be handled thoughtfully and with appropriate safeguards.

\textbf{Ethical and Methodological Considerations:}
The use of LLMs in psychology raises several important ethical and methodological concerns. One issue is their limited representation of psychological diversity, as biases in training data may prevent them from adequately capturing the range of global psychological experiences. Another concern is their opaque mode of operation, since the black-box nature of these systems makes it difficult to understand how outputs are generated, thereby reducing interpretability and trust. There is also the broader risk that over-reliance on LLMs may weaken human agency while reproducing or amplifying biases present in training data \cite{132,133}. 

In response to these concerns, researchers have called for greater transparency in how LLMs are used and evaluated, the development of standardized benchmarks for assessing performance in psychology, the creation of high-quality and diverse keystone datasets to improve generalizability, and the expansion of shared infrastructure that can support collaboration across disciplines in addressing both ethical and technical challenges \cite{132}.

In conclusion, while LLMs are not yet ready for transformative applications, researchers propose actionable steps to address current limitations. These include creating robust datasets, standardizing evaluation methods, and promoting interdisciplinary collaboration to refine LLM capabilities and applications. These efforts are expected to unlock new possibilities in psychological measurement, experimentation, and practice \cite{133}.

\subsubsection{Economics}
The advent of Large Language Models (LLMs) like GPT has introduced transformative opportunities in economics, enabling applications across various domains. This section examines their impact, organized into three key themes: productivity enhancement, applications in economic decision-making and prediction, and limitations in economic reasoning and knowledge comprehension.

\textbf{Productivity Enhancement:}
LLMs have demonstrated significant potential in enhancing productivity within economics research. By automating micro-tasks such as ideation, writing, and data analysis, LLMs streamline research processes, allowing economists to focus on higher-order creative and analytical tasks. Notable applications include generating content for blogs, presentations, interviews, and podcasts aimed at disseminating research findings \cite{134,135,136}. 

Additionally, LLMs provide interactive workspaces, improve reasoning capabilities, and offer enhanced internet search functionalities. These features position LLMs as essential tools for increasing efficiency and reducing the time investment required for routine tasks, enabling researchers to focus on innovation and deeper inquiry.

\textbf{Application in Economic Decision-Making and Prediction:}
Beyond their role in enhancing productivity, LLMs have also shown promise in economic decision-making and predictive applications. They can be used to generate training data for complex economic experiments, including language-based persuasion games designed to simulate human decision-making \cite{137}. In addition, their language comprehension and generative abilities allow them to support behavioral modeling by helping design experiments that approximate real-world economic behavior and provide insight into the processes underlying human choice and decision-making. These applications point to the growing potential of LLMs to assist economists in modeling and analyzing complex economic settings, thereby opening new avenues for studying decision-making in both theoretical and applied contexts.

\textbf{Limitations in Economic Reasoning and Decision-Making:}
Despite these strengths, LLMs continue to face important limitations in reasoning and comprehension within the economic domain. A recent study using the EconNLI dataset examined their ability to reason about economic events and interpret economic knowledge, and the findings suggest that these models often generate incorrect or fabricated outputs, especially in tasks that require nuanced understanding and careful judgment \cite{138}. 

Such limitations highlight the risks of relying on LLMs in high-stakes economic analysis and decision-making, where accuracy and contextual sensitivity are essential. Their performance is further constrained by a lack of domain-specific expertise, since general-purpose training leaves important gaps in economic knowledge, and by their tendency to hallucinate, producing plausible but inaccurate information that further weakens their reliability in sensitive settings.

\subsubsection{Political Science}
The advent of Large Language Models (LLMs) has introduced transformative changes to political science research, offering both new opportunities and notable challenges. Recent literature underscores their potential to transform the field by facilitating the exploration of novel research areas and advancing political methodologies.

LLMs, particularly tools like GPT, are reshaping traditional research practices. This automation saves researchers time and money, as data collection and analysis are far more efficient \cite{139}. Researchers can now extract information from diverse and expansive data sources more easily, enhancing both the depth and reliability of their analyses. Automation thus becomes all the more invaluable for researchers with limited resources, making advanced research tools usable for a wider part of the community and enhancing overall democratic participation in empirical studies. As such, the integration of LLMs has not only expanded the efficiency and scope of political science research but also highlighted the growing importance of artificial intelligence in the discipline \cite{139}.

Beyond streamlining traditional methodologies, LLMs open up new avenues for inquiry. For instance, their ability to process and analyze large volumes of textual data enables innovative research designs that were previously infeasible. The application of LLMs in analyzing political discourse, public opinion, and policy documents illustrates their utility in tackling complex and data-intensive questions \cite{140}. However, as these models become more entrenched in research practices, scholars must grapple with ethical and methodological challenges. Issues such as data privacy, algorithmic bias, and the interpretability of AI-generated results necessitate careful consideration to ensure the responsible use of these tools \cite{140}.

In conclusion, LLMs represent a paradigm shift in political science research. Their ability to streamline data processes, democratize access to research tools, and open new research trajectories underscores their transformative potential. Nevertheless, the field must address the ethical and methodological implications accompanying their adoption to fully harness their benefits responsibly.

\subsubsection{Philosophy}
The emergence of large language models (LLMs) like GPT has sparked significant interest in their applications within philosophical research and pedagogy. The following review synthesizes recent findings on the philosophical capabilities, limitations, and educational implications of LLMs.

\textbf{Philosophical Text Generation and Perception:}
One area of research investigates the ability of LLMs to generate philosophical content that mirrors expert-level discourse. Studies suggest that models such as GPT-3 can produce texts that are comparable to those authored by professional philosophers, including figures like Daniel Dennett.

While trained experts can distinguish between AI-generated and human-written philosophical texts at a rate above chance, the distinction is less robust than anticipated. For general audiences and non-experts, identifying AI-authored responses is even more challenging. This underscores the ability of LLMs to convincingly replicate philosophical reasoning, particularly for non-specialist readers. However, the nuanced differences detected by professionals highlight the limits of LLMs in fully mimicking human philosophical expertise \cite{141}.

\textbf{Theological and Religious Texts as Benchmarks:}
A novel approach to evaluating the philosophical capabilities of LLMs involves using religious and theological texts as benchmarks. These texts, rich in complexity and interpretive demands, serve as a rigorous testing ground for assessing LLM reasoning.

Research explores how LLMs handle challenges such as interpreting the four senses of Scripture (literal, allegorical, moral, and anagogical) and engaging with abstract theological statements and deductive logic derived from religious doctrines.

By comparing models like ChatGPT, Bing, Bard, and Llama2, researchers highlight differences in their interpretive, deductive, and ontological reasoning abilities. This approach not only underscores the strengths and limitations of these systems but also establishes a structured framework for evaluating their philosophical competencies using standardized benchmarks \cite{142}.

\textbf{Pedagogical Implications:}
The integration of LLMs into philosophy education presents both challenges and opportunities. On the one hand, the ease with which students can use these models to generate seemingly well-crafted essays raises concerns about diminishing critical engagement and stunting philosophical growth. On the other hand, innovative pedagogical approaches, such as "LLM dialogues," where students engage in structured philosophical conversations with LLMs, have been proposed to enhance learning outcomes. Based on teaching experiences, this method shifts the focus from traditional essay writing to interactive, dialogical engagement, fostering critical thinking and philosophical exploration in novel ways. By leveraging the dialogic capabilities of LLMs, educators can transform potential threats into tools for enriching philosophical education \cite{143}.

In sum, LLMs represent a promising yet complex addition to philosophical research and education. While they demonstrate impressive capabilities in generating philosophical discourse and engaging with challenging benchmarks, their limitations underscore the need for careful integration into academic practices. In educational contexts, thoughtful application of LLMs, such as interactive dialogues, can enhance learning outcomes and stimulate philosophical inquiry. Future work should focus on refining LLM capabilities and addressing ethical concerns to ensure they complement, rather than compromise, human philosophical engagement.

\subsubsection{History}
The integration of Large Language Models (LLMs) into historical research has introduced transformative tools for analysis, visualization, and interpretation, complementing traditional methodologies. This review highlights advancements in LLM applications to historical studies, emphasizing their contributions to data processing, interactive research methodologies, and semantic insights in complex textual corpora.

\textbf{Enhancing Historical Text Analysis with HistoLens:}
HistoLens, a new framework built on LLM capabilities, illustrates the potential of multi-layered analysis in historical research. Using the \textit{Yantie Lun} from the Western Han dynasty as a case study, the framework applies LLMs to identify key figures, events, and concepts through named entity recognition, construct knowledge graphs that map relationships among entities, and generate geographic information visualizations that connect spatial data to historical narratives. In doing so, HistoLens enables historians to investigate particular ideologies in greater depth and uses explainable machine teaching to reveal more nuanced interpretations within historical texts. By combining quantitative analysis with visual representation, the framework improves both accessibility and understanding for researchers and educators, marking an important development in digital historical methodology \cite{144}.

\textbf{Conversational Methodologies in Historical Research:}
Recent advances in specialized vector embeddings have introduced more conversational approaches to historical inquiry. With these enhancements, LLMs allow researchers to interact dynamically with customized corpora that may include both primary source materials and secondary analyses or commentaries. This conversational mode extends beyond the capabilities of conventional search tools by enabling more nuanced question-answering and targeted data extraction suited to particular research aims. As a result, historians are now able to conduct more private and tailored investigations, even when working with archives that fall outside the models’ original training data. These developments simplify engagement with complex historical narratives and make it possible to explore diverse sources in a more intuitive way without requiring extensive technical expertise \cite{145}.

\textbf{Semantic Analysis of Historical Legal Documents:}
The semantic capabilities of LLMs have also proven especially useful in the study of historical legal documents. For example, research on digitized handwritten Greek contracts shows that models such as ChatGPT-3.5 and Gemini/Bard can effectively identify contract types by categorizing documents according to their purpose or legal function, extract relational dynamics such as landlord-tenant interactions, and recognize subjects by detecting parties and major topics within the documents.

These capabilities significantly expedite document processing, allowing for more efficient analysis of large corpora. However, LLMs exhibit limitations in recall compared to human experts, particularly when identifying subtle or ambiguous details. This highlights the importance of combining automated semantic analysis with expert validation to ensure comprehensive and accurate interpretations \cite{146}.

In summary, the application of LLMs in historical research represents a paradigm shift, enabling richer exploration and analysis of historical texts and narratives. Frameworks like HistoLens, conversational research methodologies, and advancements in semantic processing demonstrate the transformative potential of these models. Nonetheless, human expertise remains essential to address their limitations and ensure that automated insights are both valid and nuanced. Future work should focus on refining these tools, increasing their scope, and encouraging interdisciplinary collaboration in order to maximize their impact on historical scholarship.

\subsubsection{Linguistics}
Large Language Models (LLMs) have become powerful tools of linguistics research. Their ability to process and generate natural language text at scale offers unprecedented opportunities for enhancing expert productivity. One significant application of LLMs lies in their capacity to assist in validating or refuting linguistic judgments. This capability is particularly valuable in areas where human expertise is traditionally required for detailed classification and annotation tasks.

This is demonstrated through two zero-shot prompting experiments: one on selecting and annotating lexical units related to FrameNet semantic frames, and another on categorizing verbs into aspectual classes, which are linked to semantic frame evocation. The study highlights the integration of linguistic theories, the design of prompts, and the outcomes of these experiments to show LLMs' utility in linguistic classification and annotation tasks \cite{147}.

\subsubsection{Archaeology}
Generative AI—particularly Large Language Models (LLMs), including the model ChatGPT—has been transformative in archaeological research. In recent years, these technologies have become far more accessible, allowing researchers and practitioners without programming expertise to use AI to generate novel applications. For instance, ChatGPT has been employed to support the development of complex tools for analyzing aerial and satellite imagery to detect archaeological anomalies. This functionality not only democratizes technological engagement but also facilitates interdisciplinary collaboration by bridging gaps between the humanities and computer science \cite{148}.

The potential impact of generative AI extends beyond technical applications to include natural language interactions and content generation. This capability makes LLMs highly versatile for various archaeological practices, from research design to data interpretation. However, while the accessibility and functionality of LLMs are promising, their meaningful integration into the field presents challenges. These include ensuring accuracy, addressing biases, and refining application strategies to align with archaeological methodologies. Nevertheless, as these technologies evolve, their potential to reshape research practices and expand the methodological toolkit of archaeologists is considerable \cite{149}.

\subsubsection{Geography}
The advent of large language models (LLMs), such as GPT, has begun to reshape the landscape of geographic research and education. These models hold significant potential for enhancing curriculum development and advancing geospatial science through innovative applications. However, their integration into academic practices is not without challenges, underscoring the necessity for geographers to develop AI literacy. This foundational skill enables researchers to navigate the ethical concerns and limitations inherent in LLM use, ensuring responsible and effective engagement with these technologies \cite{150}.

A systematic review of the role of LLMs in geospatial science further emphasizes their transformative impact on the field. By analyzing 26 studies, researchers identified the diverse tasks and data types associated with LLMs and generative AI models, highlighting their ability to address complex geospatial challenges. Despite these advancements, significant obstacles remain, such as data biases, transparency issues, and the need for tailored research frameworks. The review also outlines future research directions to guide geographers in leveraging LLMs effectively while fostering innovation and maintaining a competitive edge in evolving scientific paradigms \cite{151}.

\begin{figure}[tp]            
    \centering
    \includegraphics[
        width=1.0\textwidth,    
        height=0.88\textheight, 
        keepaspectratio
    ]{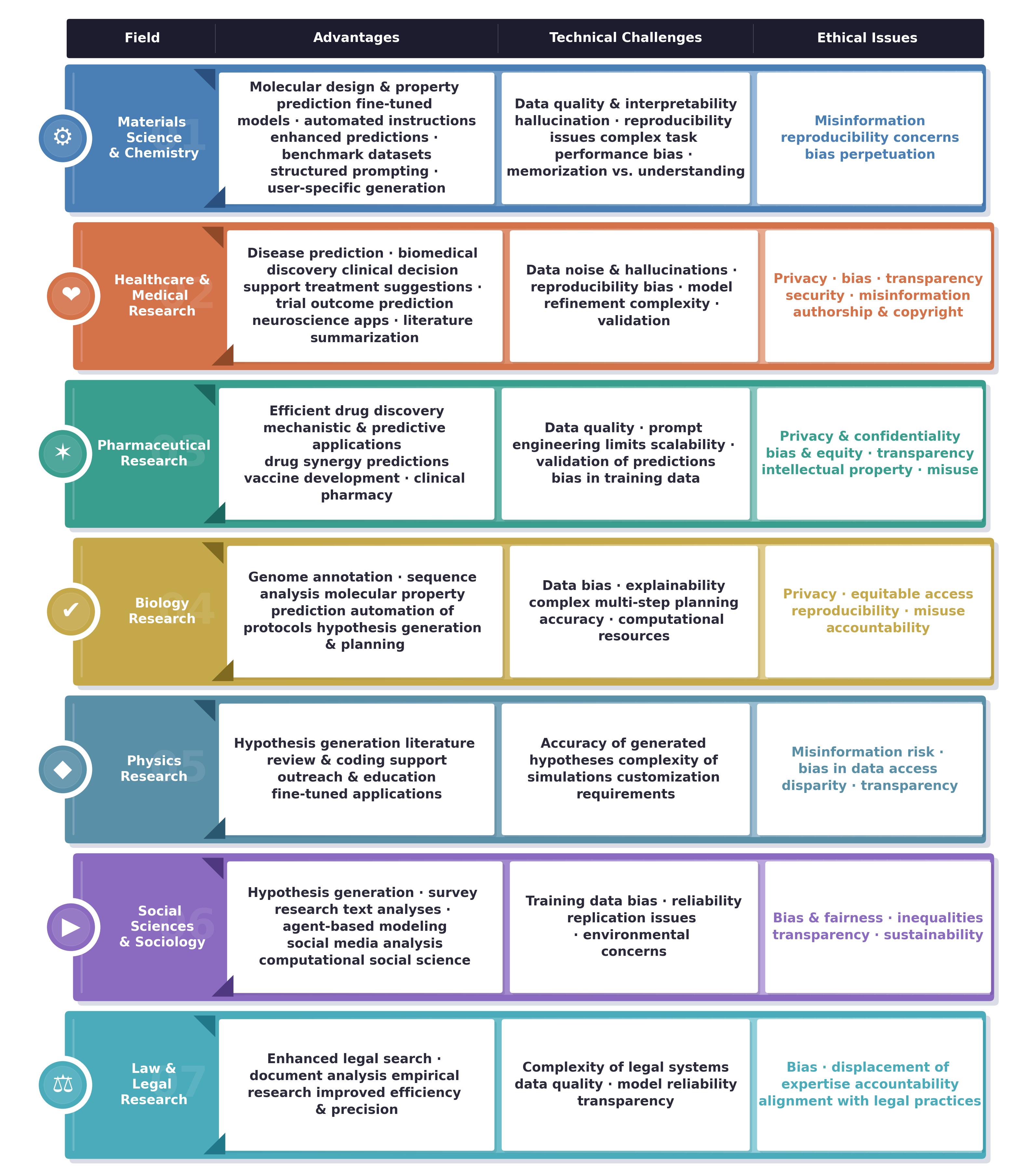}
    \caption{Opportunities, Technical Challenges, and Ethical Issues
             Related to LLM Use Across Research Fields.}
    \label{fig:table3_full_science}
\end{figure}

\begin{figure}[tp]            
    \centering
    \includegraphics[
        width=1.0\textwidth,    
        height=0.88\textheight, 
        keepaspectratio
    ]{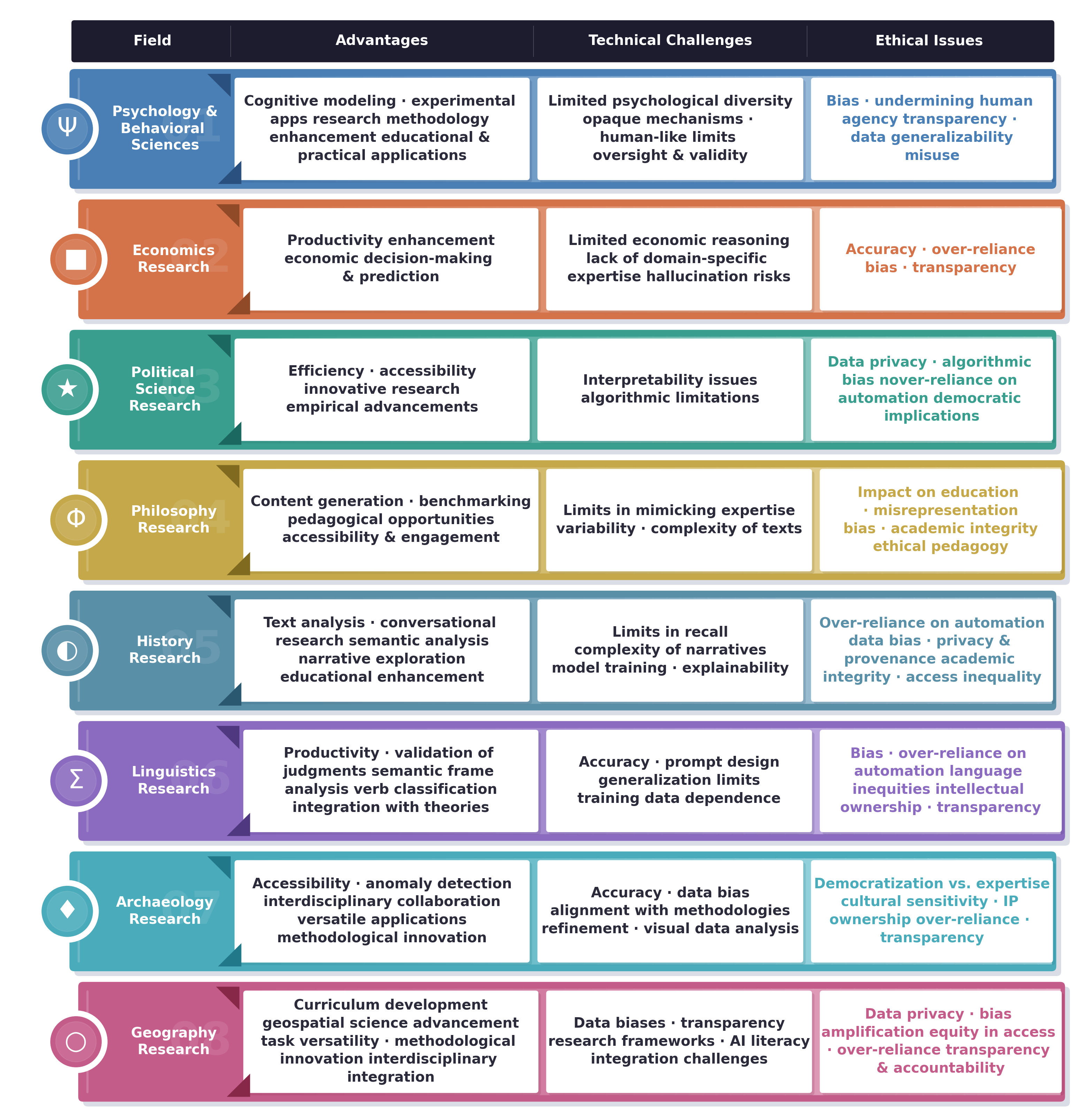}
    \caption{Opportunities, Technical Challenges, and Ethical Issues
             Related to LLM Use Across Research Fields. (continued)}
    \label{fig:table3_full_science}
\end{figure}

\section{Discussion}
The preceding survey of LLM applications across the natural sciences, social sciences, and humanities reveals a landscape marked by both significant opportunity and persistent challenge. As these models become more deeply integrated into research workflows, it is important to move beyond individual field-specific observations and consider the broader patterns that emerge across disciplines. The following discussion synthesizes the evidence gathered in this review to offer a comparative perspective on how LLMs are being received across fields, to identify the advantages and limitations that recur most consistently, and to examine the ethical dimensions that cut across disciplinary boundaries. Together, these considerations point toward the conditions under which the scientific community can most responsibly and effectively harness the transformative potential of large language models.

\subsection{Reception of LLMs in Various Fields: A Comparative Perspective}
Large Language Models (LLMs) are used across the natural sciences and the humanities/social sciences, although their application is more widespread in the natural sciences due to the technical and diverse challenges they address. In these fields, LLMs have been particularly transformative in domains like medicine and healthcare, where they assist in clinical decision-making, disease progression analysis, and drug discovery. Additionally, LLMs are making notable contributions to chemistry and materials science, where fine-tuned models predict molecular properties and streamline research processes. Conversely, in social sciences and humanities, LLMs excel in sociology, computational social science, and legal studies, supporting tasks such as hypothesis generation, legal document analysis, and simulating social phenomena.

LLMs can perform many tasks exceedingly well: data analysis and synthesis, predictive modeling, scientific writing, and fostering creativity. In healthcare, for instance, LLMs provide support with the analysis of medical records, summarization of clinical trial results, and the prediction of disease course. Similarly, in law, LLMs are employed for contract analysis and legal research, helping to streamline administrative and legal processes. These advantages come with challenges, however: in the natural sciences, there are concerns around model reproducibility, interpretability, and integration with legacy knowledge; in social science, the possibility of biased or unethical results is at the forefront of considerations. The ethical dimensions of data privacy and bias—more than any other issues—remain the most important gatekeepers to further adoption across medicine and law.

In summary, LLMs are most widely used in the natural sciences, especially within healthcare, medicine, and materials science. Their applications in the social sciences and humanities are increasing but with more caution, with the ethical implications of bias and fairness and misuse of AI technologies being a topic of much debate. As these LLMs continue to evolve, so too will the challenges they also bring with them in matters of accuracy, interpretability, and ethics, which will always be at the forefront in their integration into scientific research and practice.

\subsection{Advantages of LLM Use in Various Fields}
A prominent advantage of large language models (LLMs) lies in their capacity to enhance productivity, foster novel applications, and support hypothesis generation across diverse disciplines. In the natural sciences, such as biology, chemistry, and physics, LLMs contribute significantly by automating processes like molecular property prediction and genome annotation, thereby expediting scientific discovery. In the social sciences, including economics, political science, and sociology, LLMs are highly valued for their ability to process large datasets, generate hypotheses, and improve decision-making, which in turn enhances research and policy development.

In healthcare and pharmaceutical research, LLMs are particularly advantageous in clinical decision support, drug discovery, and vaccine development, offering tangible benefits in improving patient outcomes and accelerating medical advancements. In linguistics, LLMs facilitate the validation of linguistic judgments and the analysis of semantic frames, underscoring their importance in both theoretical and applied linguistic research.

In fields such as law and history, LLMs support document analysis and conversational methodologies, providing valuable tools for enhancing research efficiency. Philosophy and psychology benefit from LLMs in content generation and cognitive modeling, respectively, further expanding the scope of their application in these disciplines. Interdisciplinary fields like archaeology and geography also utilize LLMs for tasks such as anomaly detection and geospatial analysis, reflecting their broad utility.

Furthermore, LLMs are instrumental in enhancing the accessibility of fields such as philosophy and history by enabling content summarization and facilitating engagement with broader audiences. The versatility of LLMs is also evident in their role in advancing computational social science and digital media analysis within the social sciences, particularly sociology, highlighting the adaptability of these models across various domains of inquiry.

\subsection{Challenges of LLM Use in Various Fields}
Several technical challenges persist across various disciplines, including data bias, reproducibility issues, and limitations in model interpretability. These challenges are particularly prominent in fields such as psychology, law, and the social sciences, where human-like reasoning and interpretative rigor are essential. The reliance on high-quality training data is a recurring issue in healthcare, economics, and pharmaceutical research, where the real-world implications of these fields demand high levels of precision and accuracy.

In physics, the complexity of simulations and hypothesis generation presents unique obstacles, reflecting the difficulty of applying large language models (LLMs) to mathematically intensive domains. Similarly, linguistics encounters challenges related to prompt design and the limits of generalization, which underscore the intricate nature of language processing tasks. In fields like geography and archaeology, the integration of LLMs with traditional research methodologies poses alignment issues, highlighting the need for domain-specific AI literacy and tailored applications.

Biology faces significant challenges due to the computational resource requirements for multi-step planning, a consequence of the field's reliance on large datasets and intricate analyses. In disciplines such as philosophy and history, concerns about the variability of LLM performance and the complexity of generating meaningful interpretations of texts can hinder academic progress and limit the utility of AI-generated insights.

While the natural sciences, including chemistry, biology, and physics, excel in utilizing LLMs for predictive and computational tasks, they continue to face challenges related to the alignment of AI-generated insights with experimental data and the significant resource requirements associated with these applications. In healthcare and pharmaceutical research, LLMs offer notable benefits, particularly in clinical decision support and treatment development, but ethical considerations, such as privacy concerns and the need for validation, remain critical.

In the social sciences, including sociology, LLMs have proven valuable for innovative research on social behaviors; however, challenges related to data reliability and the potential exacerbation of societal inequalities must be carefully addressed. In the humanities, particularly philosophy and history, LLMs have the potential to broaden access to knowledge and democratize scholarly engagement. However, ethical risks, such as threats to academic integrity and the potential misuse of AI-generated content, require vigilant oversight and careful management.

\subsection{Ethical Challenges of LLM Use in Various Fields}
Ethical concerns such as bias, misinformation, transparency, and privacy cut across a wide range of academic disciplines. These concerns are not uniform but are instead closely tied to the particular values and methodological demands of each field. Bias, for instance, remains a serious issue in areas such as the social sciences, law, and healthcare, where fairness, equity, and impartiality are central. In such contexts, the possibility that AI systems may reproduce or reinforce existing structural biases threatens both the credibility of research findings and the fairness of policy or clinical decisions. Privacy is likewise especially important in healthcare, pharmaceutical research, and political science, where researchers often work with sensitive personal or political information. In these domains, the use of LLMs raises pressing questions about consent, confidentiality, data security, and the potential misuse of private information.

In the fields of history and philosophy, ethical challenges arise concerning academic integrity and the potential over-reliance on automation. This dependency could undermine traditional scholarly methods, leading to concerns about the devaluation of human expertise and critical thinking in these disciplines.

In archaeology, the tension between democratizing knowledge and preserving expertise highlights ethical dilemmas related to cultural sensitivity and intellectual property. The use of AI tools in this field must be navigated carefully to respect indigenous knowledge and ownership while making data more accessible.

Economics and political science present ethical dilemmas regarding the over-reliance on automated systems for decision-making. The increasing use of AI in these fields raises concerns about the potential for automation to distort democratic processes, reinforce inequalities, or limit the accountability of human decision-makers.

In psychology, there are concerns about the potential to undermine human agency and the generalizability of data when relying on AI-driven models. The use of LLMs and similar technologies in psychological research must be critically examined to ensure that they do not misrepresent or over-simplify complex human behaviors.

Physics and geography bring attention to issues of access disparity and equity, particularly with regard to the global distribution of technological resources and benefits. These fields raise important questions about the inclusivity of AI-driven advancements and the potential for widening the gap between regions and populations with varying levels of access to cutting-edge technologies.

Each of these ethical concerns underscores the need for careful consideration and responsible stewardship in the application of AI technologies across disciplines, ensuring that the benefits of these tools are equitably distributed while minimizing harm.

\subsection{Additional Ethical Challenges of LLM Use in Scientific Research}
The integration of LLMs into scientific research carries transformative potential but also introduces a spectrum of ethical challenges. While some of these issues are recognized in the literature, others remain under-explored, bringing into question the integrity, autonomy, and equity of scientific inquiry. This article explores ten emerging ethical challenges and proposes pathways for addressing them.

\textbf{Loss of Human Agency in Research Design:}
The increasing reliance on LLMs for tasks traditionally requiring human creativity—such as hypothesis generation, experimental design, and data interpretation—poses risks to researcher autonomy. This over-dependence could shift scientific priorities toward AI-driven agendas, narrowing intellectual diversity and diminishing exploratory research. For instance, in fields like behavioral sciences, AI-designed studies may prioritize correlations over causations, limiting opportunities for innovative breakthroughs. Ensuring a balanced integration of human insight and machine intelligence is essential to preserving the integrity of scientific exploration.

\textbf{AI-Driven Confirmation Bias in Research:}
LLMs, trained on existing datasets, may inadvertently reinforce dominant paradigms while overlooking unconventional or minority perspectives. This "AI-driven confirmation bias" risks amplifying entrenched theories, thereby stifling innovation. For example, reliance on LLM-generated literature reviews could marginalize underrepresented scientific voices. Addressing this challenge requires diversifying training datasets and incorporating mechanisms for promoting epistemic diversity in AI-driven research tools.

\textbf{Over-Reliance on Black Box Models:}
LLMs’ opacity—often referred to as their "black box" nature—raises serious concerns about accountability in high-stakes research domains like medicine or environmental science. For example, an LLM used in drug discovery may suggest compounds without clear explanations of its reasoning, posing risks if the outputs are accepted uncritically. Researchers must adopt explainable AI (XAI) methodologies and robust validation protocols to ensure transparency and reliability.

\textbf{Ethical Implications of AI-Generated Co-Authorship:}
The question of whether LLMs should be credited as co-authors introduces complexities surrounding intellectual ownership and accountability. For example, an LLM that drafts sections of a research paper may blur the lines between tool and collaborator. Drawing on existing authorship guidelines, such as those proposed by COPE, it is crucial to establish ethical frameworks that recognize AI’s contributions without compromising the value of human authorship.

\textbf{Manipulation of Research Outcomes:}
LLMs could be misused to fabricate or skew research outcomes, particularly in areas tied to funding or regulatory approvals. For instance, fine-tuning an LLM to align with desired conclusions in a clinical trial could lead to significant ethical violations. Preventing such misuse requires robust oversight mechanisms, transparent reporting of AI methodologies, and accountability frameworks to ensure scientific integrity.

\textbf{Informed Consent and Data Use:}
LLMs are normally trained on vast amounts of data that include private or sensitive information. The training of an LLM using unpublished clinical trial data without explicit authorization is contrary to ethical principles related to informed consent and intellectual property rights. Calls for clearer guidelines on regulating data use will make sure transparency, consent, and accountability principles are followed in AI-driven research by researchers.

\textbf{The Impact on Research Funding and Employment:}
Automation of research tasks may displace early-career researchers and other academic professionals, including literature reviews, data analysis, and even scientific writing. But it will also enable possibilities for reskilling and reimagining roles within academia. Institutions and funding agencies must proactively develop reskilling programs and equitable funding policies to ensure that the efficiency gains from LLM adoption benefit the broader research community.

\textbf{Unequal Access to LLMs:}
Access to LLMs remains concentrated in well-funded institutions and private corporations, exacerbating global inequities in scientific research. Researchers in low-income regions or smaller institutions may face significant barriers to accessing cutting-edge AI tools. Addressing this disparity requires fostering international collaborations, supporting open-access initiatives, and advocating for equitable distribution of AI resources.

\textbf{AI-Driven Ethical Dilemmas in Research Conduct:}
LLMs trained on historical datasets may perpetuate outdated or unethical practices. For example, an LLM trained on pre-2000 studies in medical research may recommend experimental designs now deemed unethical. To prevent such outcomes, researchers must critically assess AI-generated recommendations and integrate ethical review processes into LLM-driven workflows.

\textbf{Ethical Implications of AI-Generated Scientific Impact:}
AI-generated outputs may be misrepresented as groundbreaking contributions, inflating perceptions of scientific progress. For instance, publishers could showcase AI-drafted papers without transparent attribution, undermining academic integrity. Acknowledging the role of AI transparently and ethically is essential to maintaining trust in scientific practices and ensuring that human contributions are not overshadowed.

\begin{figure}[H]
    \centering
    \includegraphics[
        width=0.9\textwidth,
        height=0.55\textheight
    ]{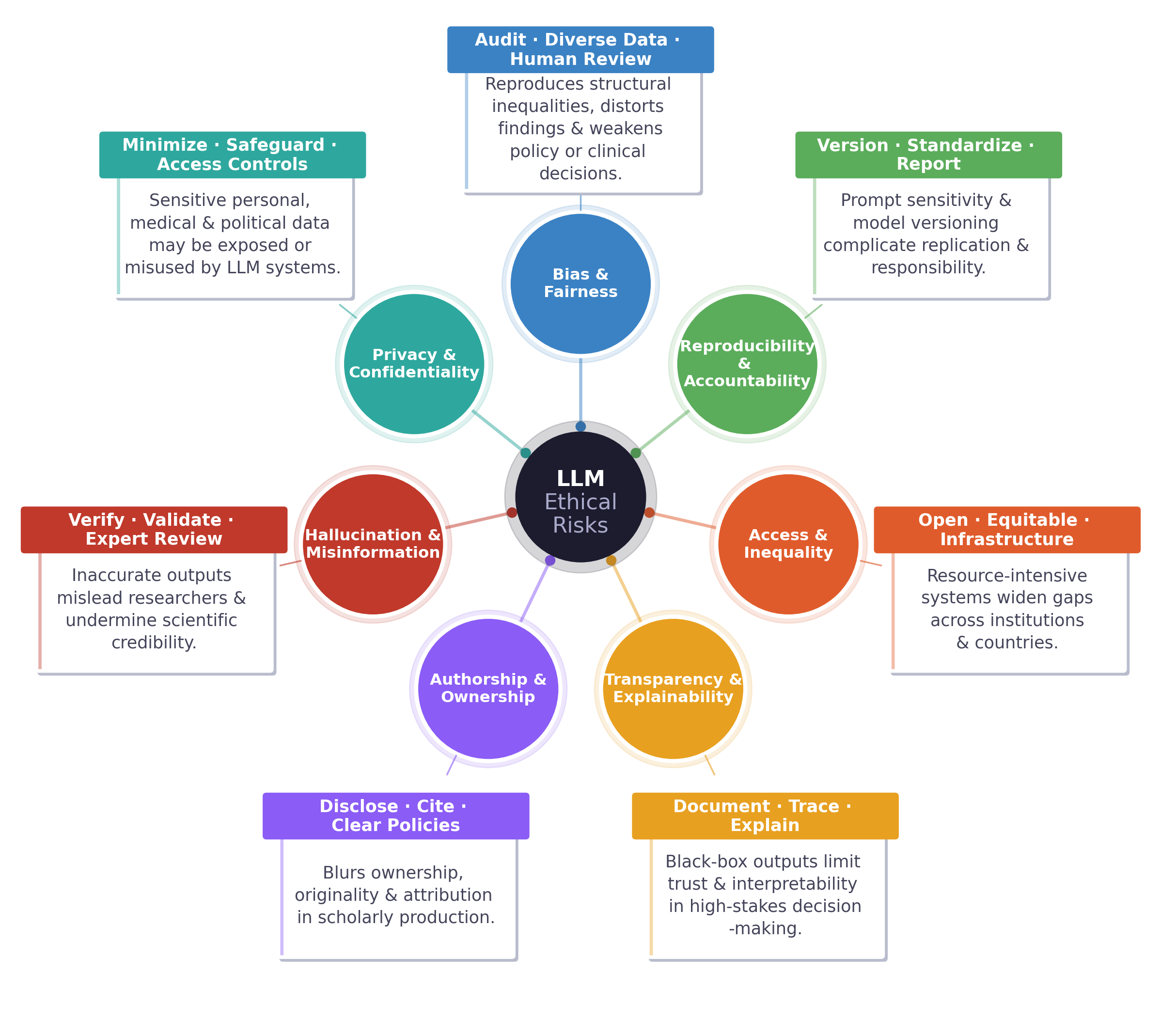}
    \caption{Cross-Field Ethical Risks and Governance Priorities for LLM-Based Research
}
    \label{fig:ethics_network}
\end{figure}

In sum, LLMs offer transformative potential for scientific research, but their ethical integration requires careful consideration of challenges ranging from researcher autonomy to global equity. Such a framework would require collaboration between researchers, policymakers, and AI developers in establishing a framework that holds the principles of integrity, accountability, and inclusivity. Only an interdisciplinary approach to AI ethics will allow the scientific community to guarantee that the adoption of LLMs will only serve to enrich, rather than compromise, the quest for knowledge.

\section{Conclusion and Future Directions}
Large Language Models have become a transformative part of scientific research, spanning the natural sciences to social science and humanities. They demonstrate a remarkable capacity to process huge data amounts, formulate hypotheses, and assist with scientific writing or modeling of complex systems for many different research contexts. However, their growing use also brings up a few ethical and technical problems due to increasing reliance on LLMs. These include the potential erosion of human agency in research design, reinforcement of confirmation biases, and the model interpretability problems commonly referred to as the "black box" problem. These are compounded by issues of intellectual property, data privacy, and equitable access to AI technologies. As the scientific community continues to explore the potential of LLMs, it must also address these challenges through continued dialogue and collaboration.

Addressing these concerns will require several important developments. First, interdisciplinary collaboration is essential for improving the capabilities and performance of LLMs. Bringing together researchers from computer science, linguistics, ethics, and domain-specific fields such as medicine, law, and materials science can support the fine-tuning of models so that they better reflect the needs and constraints of different areas of research. Such collaboration can also contribute to the development of more robust, context-sensitive, and adaptable systems for domain-specific problems. Second, greater standardization and benchmarking are needed if LLMs are to be used reliably in scientific research. The creation of standardized datasets and evaluation frameworks, as seen in initiatives such as TextEdge for materials science, provides a useful path for assessing model performance in specialized contexts. More consistent benchmarks and metrics would improve model training and make it easier to evaluate reproducibility and comparability across studies. Third, the expansion of ethical frameworks is necessary to guide the responsible use of LLMs in research. These frameworks should directly address issues such as bias, misuse, inequality, privacy, intellectual property, and human dignity, while also evolving in response to new technological developments so that transparency and accountability remain central. Finally, improving explainability will be crucial, particularly in high-stakes domains such as healthcare, law, and drug discovery, where the black-box character of LLMs continues to limit trust. Strengthening interpretability would make it easier to understand how models arrive at particular outputs and would help ensure that decisions informed by these systems remain grounded in sound scientific reasoning. In areas where accountability and transparency are essential, greater explainability will be a necessary condition for broader adoption.

By addressing these areas, the scientific community can better navigate the challenges posed by LLMs while unlocking their full potential. The continued development of more accurate, transparent, and ethically grounded LLMs will shape the future of scientific research, fostering more efficient, equitable, and innovative approaches to knowledge generation.

\section*{Acknowledgement}

 In accordance with MLA (Modern Language Association) guidelines, we acknowledge the use of AI-powered tools, such as Anthropic's applications, for their assistance and support including in writing, editing, and improving the clarity and organization of the manuscript

\section*{Competing interests}
The authors declare that the research was conducted in the absence of any commercial or financial relationships that could be construed as a potential conflict of interest.

{\fontsize{8}{10}\selectfont
\bibliography{main}
}

\end{document}